\documentclass[aps,prl,reprint,superscriptaddress,showpacs]{revtex4-1}
\usepackage{color,graphicx,calc,url}
\usepackage{amsmath}
\usepackage{bm}
\usepackage{dcolumn}
\usepackage{amsfonts}
\usepackage{amssymb}
\usepackage{wasysym}
\usepackage{mathrsfs} 
\usepackage{subfig}
\usepackage{gensymb}
\usepackage[ppl]{mathcomp}
\usepackage[breaklinks=true]{hyperref}
\hypersetup{colorlinks=true, citecolor=blue, filecolor=blue, linkcolor=blue, urlcolor=blue}

\begin{document}
	
	\title{Negative spin Hall magnetoresistance of Pt on the bulk easy-plane antiferromagnet NiO}
	

	\author{Geert R. Hoogeboom}
	\email[]{g.r.hoogeboom@rug.nl}
	\author{Aisha Aqeel}
	\email[]{Present adress: Institute of Experimental and Applied Physics, University of Regensburg, Universit\"atsstr. 31, 93053 Regensburg, Germany}
	\author{Timo Kuschel}
	\author{Thomas T.M. Palstra}
	\email[]{Present adress: University of Twente, Drienerlolaan 5, 7522 NB Enschede, The Netherlands}
	\author{Bart J. van Wees}
	\affiliation{Physics of Nanodevices, Zernike Institute for Advanced Materials, University of Groningen, Nijenborgh 4, 9747 AG Groningen, The Netherlands}
	\date{\today}
	
	\begin{abstract}
		
		We report on spin Hall magnetoresistance (SMR) measurements of Pt Hall bars on the antiferromagnetic NiO(111) single crystal. An SMR with a sign opposite of conventional SMR is observed over a wide range of temperatures as well as magnetic fields stronger than 0.25T. The negative sign of the SMR can be explained by the alignment of magnetic moments being almost perpendicular to the external magnetic field within the easy plane (111) of the antiferromagnet. This correlation of magnetic moment alignment and external magnetic field direction is realized just by the easy-plane nature of the material without the need of any exchange coupling to an additional ferromagnet. The SMR signal strength decreases with increasing temperature, primarily due to the decrease in N\'eel order by including fluctuations. An increasing magnetic field increases the SMR signal strength as there are less domains and the magnetic moments are more strongly manipulated at high magnetic fields. The SMR is saturated at an applied magnetic field of $6$~T resulting in a spin-mixing conductance of $\sim10^{18}~ \Omega^{-1}$m$^{-2}$, which is comparable to that of Pt on insulating ferrimagnets such as yttrium iron garnet. An argon plasma treatment doubles the spin-mixing conductance.
		
	\end{abstract}
	
	
	\maketitle
Antiferromagnets (AFMs) are mostly known for their exchange bias pinning effect on adjacent ferromagnetic (FM) layers. Owing to the robustness against magnetic perturbations of easy-axis AFMs, this coupling allows for giant\cite{Barnas:1989, Binasch:1989} and tunnel\cite{Julliere:1975} magnetoresistance devices. More recently, metallic AFM moments have been manipulated and read out by applied spin polarized charge currents.\cite{Wadley:2016} Insulating AFMs can have spin waves carrying spin angular momentum\cite{Cheng:2014, Tserkovnyak:2014, Xiao:2015} which switch ultrafast\cite{Rasing:2004} and act as efficient spin-current transmitter,\cite{Tserkovnyak:2014, Brataas:2016} thus, playing an important role in spintronic applications.\cite{Jungwirth:2016}

Injection and detection of spin angular momentum in insulating magnets can be done by the combination of the spin Hall effect (SHE)\cite{Kato:2004} and the inverse spin Hall effect (ISHE)\cite{Tatara:2006} in normal metals (NMs). Rotating  the magnetic moments in the adjacent magnet by an applied magnetic field can change the interaction of the Pt spins with the magnet. This leads to spin Hall magnetoresistance (SMR)\cite{Saitoh:2013, Vlietstra:2013, Bauer:2013, Goennenwein:2013} which enables the study of various magnetic systems. In collinear ferrimagnets\cite{Saitoh:2013, Vlietstra:2013, Bauer:2013, Goennenwein:2013}, the magnetic moments align collinear to the external magnetic field, resulting in positive SMR contributions. A reversed angular modulation, or negative SMR signal has been observed when the average canting angle between the magnetic moments and the external magnetic field exceeds 45$\degree$ in canted magnetic systems\cite{Ganzhorn:2016} The localized spins of bulk AFM´s in an easy plane are nearly perpendicular to the applied magnetic field.\cite{Kurosawa:1980, Roth:1960, Slack:1960} The perpendicular alignment is expected to create a negative SMR due to the $90\degree$ angle shift, but this effect has not yet been studied in detail.

Spin-transfer measurements through insulating AFMs have been studied with stacked Pt/NiO/YIG devices. Magnons are created in YIG (yttrium iron garnet, Y$_3$Fe$_5$O$_{12}$) by ferromagnetic resonance\cite{Yang:2014, Kent:2016, QiuSaitoh:2016} or the spin Seebeck effect,\cite{Heremans:2016, Tokura:2015, Chien:2016} propagate through the NiO layer and are detected in Pt by the ISHE. For NiO layers thicker than $\sim$5~nm, the transmitted spin current decreased rapidly with thickness. The sign of the SMR signal in these Pt/NiO/YIG stacks is observed to be positive at room temperature and  becomes negative at low temperatures.\cite{Saitoh:2016, Chienelectrical:2016, Shang:2016} The authors explain this domination of the positive SMR at room temperature by spin currents injected at the Pt/NiO interface, transmitted through NiO and partly reflected when entering the YIG. At low temperatures, the spin currents towards and from the YIG are suppressed due to the vanishing spin transmittance in NiO, thus, the total signal is dominated by the negative SMR from NiO. For the Pt/NiO/YIG samples, the NiO magnetic moments are indirectly aligned perpendicular to the magnetic field via an exchange coupling 
with YIG which is saturated at 0.06~mT. \cite{Saitoh:2016, Chienelectrical:2016, Shang:2016}\

In this letter, SMR signals are obtained from Pt/NiO heterostructures by direct manipulation of the AFM spins in the magnetic easy plane with an applied magnetic field and without the need of any exchange coupling to an additional ferro- or ferrimagnet. The surface of the studied NiO bulk single crystal has a (111) cut, so that the Pt/NiO interface has the easy plane of the NiO magnet. A strong magnetic field will align the moments perpendicular to the magnetic field direction due to Zeeman energy reduction, aside from contributions due to magnetic anisotropy or domain formation by magnetostriction. Therefore, by rotating the magnetic field, the magnetic moments follow the rotation with almost a $90\degree$ angle shift within this magnetic easy plane.\cite{Roth:1960, Slack:1960}

	\begin{figure}
		\includegraphics[width=8.5cm]{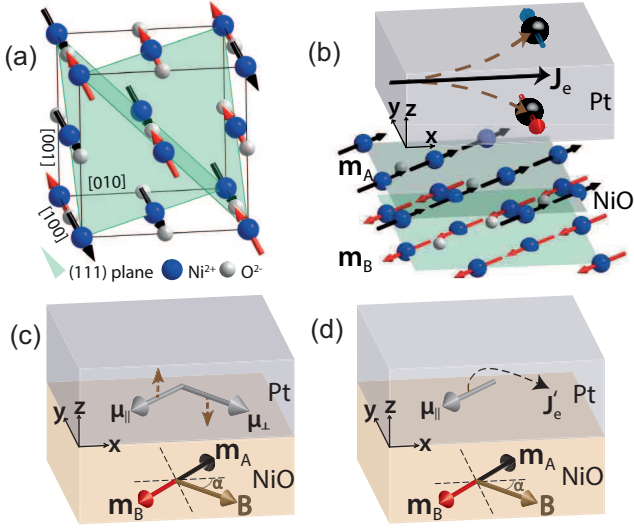}
		\caption{(a) Atomic unit cell of the NiO crystal. The blue and white balls represent Ni$^{2+}$ and O$^{2-}$ ions and the green planes indicate the (111) planes in which the moments are aligned. (b) Heterostructure of the upper (111) planes of the NiO and the Pt layer where the SHE converts lateral charge current $\mathbf{J}_e$ to vertical spin current. (c) The spin accumulation is decomposed into two components: a collinear component $\bm{\mu}_{||}$ which is reflected back into Pt and a perpendicular component $\bm{\mu}_\perp$ which is transferred to the NiO. Here, the angle $\alpha$ of the magnetic field \textbf{B} is defined with respect to the direction of \textbf{J}$_e$. (d) The spin current from the reflected collinear component is converted back into a charge current \textbf{J'}$_e$ by the ISHE. 
				}
		\label{fig:NiO_and_SMR}
	\end{figure}

		
	Figure \ref{fig:NiO_and_SMR} (a) shows the atomic face centered cubic unit cell of NiO. The superexchange interaction between Ni$^{2+}$ ions mediated by O$^{2-}$ ions aligns the Ni$^{2+}$ magnetic moments antiparallel. Below the N\'{e}el temperature of 523~K, the total interaction cause the spins to have their preferential orientation in one of the \{111\} planes. Magnetostriction creates rhombohedral distortions in the diagonal $<$111$>$ directions and causes the emergence of domains in a single crystal.\cite{Roth:1960} 
	
	In a NM with large spin-orbit coupling, the electrons deflect in a direction depending on their spin orientation, resulting in a spin current perpendicular to the charge current direction~-~known as SHE. Since NiO is an insulator, a vertical spin polarized charge current in the NM results in a spin accumulation at the interface, which is shown in Fig. \ref{fig:NiO_and_SMR}(b).  However, the spin angular momentum can be transferred to NiO when the magnetic moments of NiO are perpendicularly aligned to the accumulated spins. Figure \ref{fig:NiO_and_SMR}(c) shows the spin transfer of perpendicular ($\bm{\mu}_\perp$) and the reflection of collinear ($\bm{\mu}_{||}$) components of the spin accumulation. \
	
	The ISHE converts the reflected collinear component into charge current as shown in Fig. \ref{fig:NiO_and_SMR}(d). The spin transfer depends on the microscopic interaction of the spin accumulation with the NiO and can reduce the back-flow of the spin current depending on the direction of the magnetic moments of NiO at the interface. The changes in reflected spin current and, thus, in resistivity of the NM, can be measured both longitudinally and transversally.
	
	The spin transfer through the interface is given by \cite{Tserkovnyak:2014}
	\begin{equation}
	\mathbf{J}_s=\frac{G_r}{4\pi}\mathbf{n}\times\bm{\mu}\times\mathbf{n}+\frac{G_i}{4\pi}\bm{\mu}\times\mathbf{n}
	\label{eq:Js}
	\end{equation}
	where $\mathbf{n}$ is the N\'eel vector and $G_r$ and $G_i$ are the real and imaginary components of the spin-mixing conductance $G_{\uparrow\downarrow}$, respectively. The first part of Eq. \ref{eq:Js} containing $G_r$ is governed by Umklapp reflections and the second part containing $G_i$ is associated with specular reflections similar to the ferrimagnetic case.\cite{Tserkovnyak:2014,Bauer:2000}
	
	The exchange approximation is used since the net magnetization $\mathbf{m}=(\mathbf{m}_A+\mathbf{m}_B)/2$ is considerably smaller than $\mathbf{n}=(\mathbf{m}_A-\mathbf{m}_B)/2$, where \textbf{m}$_A$ and \textbf{m}$_B$ the magnetization of the two sublattices. Still, a small canting of the moments lowers the Zeeman energy and aligns the NiO magnetic moments almost perpendicular to the applied in-plane magnetic field. Therefore, the magnetic field couples to this small magnetization and the magnetic moment directions are following the magnetic field with about $90\degree$ angle shift.
	
	Resulting changes in the longitudinal and transverse SMR ($\rho_{L}$ and $\rho_{T}$, respectively) can be described by the magnetization of the ferromagnetic sublattices. The regular ferromagnetic SMR equations\cite{Bauer:2013, Vlietstra:2013} are adapted to
	\begin{equation}
	\rho_{L}=\sum\limits_{i=A,B}<1-\mathbf{m}_{y,i}^2>\Delta\rho_{1} + \rho+\Delta\rho_{0} 
	\label{rhol}
	\end{equation}
	\begin{equation}
	\rho_{T}=\sum\limits_{i=A,B}<\mathbf{m}_{x,i}\mathbf{m}_{y,i}>\Delta\rho_{1} +\Delta\rho_{\rm Hall}B_z
	\label{rhot}
	\end{equation}
	with $\rho$ being the electrical resistivity of Pt, $\Delta\rho_{\rm Hall}B_z$ describes the change in resistivity caused by the ordinary Hall effect with an out-of-plane component of the magnetic field $B_z$. $\mathbf{m}_{x,i}$ and $\mathbf{m}_{y,i}$ are the components of a sublattice magnetization in the $\mathbf{x}$ and $\mathbf{y}$ directions, respectively. Due to the quadratic dependence on \textbf{m}, the resulting resistivity change is equal for the two sublattices.  $\Delta\rho_{0}$ and $\Delta\rho_{1}$ are resistivity changes defined as\cite{Bauer:2013}
	\begin{equation}
	\frac{\Delta\rho_{0}}{\rho} =- \theta^2_{\rm SH} \frac{2\lambda}{d_N} \tanh\frac{d_N}{2\lambda}
	\end{equation}
	\begin{equation}
	\frac{\Delta\rho_{1}}{\rho} =\theta^2_{\rm SH} \frac{\lambda}{d_N} \textrm{Re} (\frac{2\lambda G_{\uparrow\downarrow} \tanh^{2}\frac{d_N}{2\lambda} }{\sigma+2\lambda G_{\uparrow\downarrow} \coth \frac{d_N}{\lambda}})
	\label{eq:SMR}
	\end{equation}
	where $\lambda$, $d_N$, $\sigma$ and $\theta_{\rm SH}$ are the spin relaxation length, thickness, bulk conductivity and the spin Hall angle in the NM, respectively. $G_{\uparrow\downarrow}$ is the spin-mixing conductance of the NM/(A)FM interface.  \\
	
	The two bulk single crystals investigated here are black colored due to vacancies with the dimensions of $5 \times 5 \times 1~$mm$^3$. For the device fabrication, the crystals are polished along the (111) surface in the line of the technique described by Aqeel ${et}$ ${al}$.\cite{Aqeel:2014} The crystals have been grinded (SiC P4000) and polished (diamond $6~\mu$m and $3~\mu$m, silica $0.04~\mu$m, AlOx $0.02~\mu$m). To remove any residuals of polishing, the samples are rinsed (hot water, ethanol), blow dried, rinsed again (propanol), and annealed at $200\degree$C. The crystallographic (111) surfaces are confirmed by x-ray diffraction and the peak-to-peak surface roughness is $0.24$~nm as determined by atomic force microscopy.\
	
	A $5$~nm thick Pt Hall bar with a $100 \times 1000~\mu$m$^2$ main bar and four, longitudinally $753~\mu$m separated, $100 \times 20~\mu$m$^2$ Pt side contacts was patterned by e-beam lithography. The Pt has been sputtered at a base pressure of $2.5 \times 10^{-7}$~mbar and a sputter pressure of $4.9\times 10^{-7}$~mbar. The etched sample was given an additional 15 second argon plasma exposure at 200~W before the Pt has been deposited to study the effect on the spin-mixing conduction between the Pt and NiO.\\
	
	\begin{figure}
		\includegraphics[width=8.5cm]{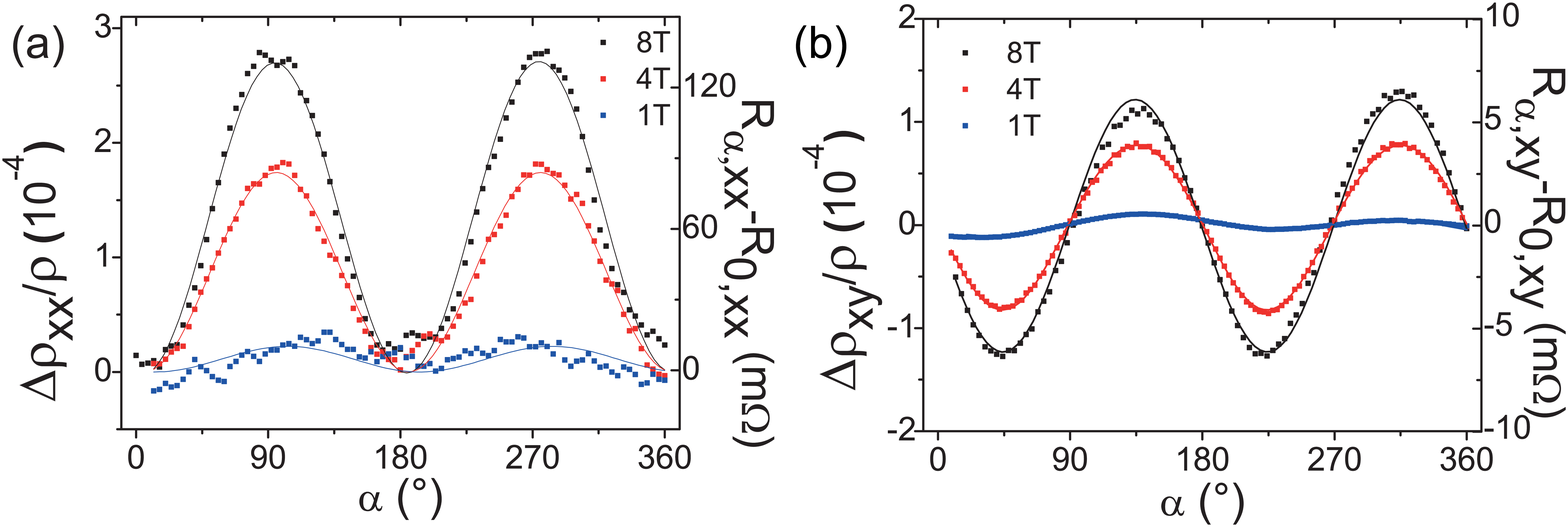}
		\caption{SMR signals in the (a) longitudinal and (b) transversal geometry, performed at 300K on the unetched sample. The right axes show the change in resistance $R_{\alpha}-R_0$ with $R_{\alpha}$ th angular dependent and $R_0$ the constant resistance; $R_{0,xx}$=460~$\Omega$ and $R_{0,xy}$=0.685~$\Omega$ for the longitudinal and transverse geometry, respectively. The left axes show the relative change in resistivity, where $\frac{\Delta\rho_{xx}}{\rho}$ ($\frac{\Delta\rho_{xy}}{\rho}$) is $\frac{R_{\alpha,xx}-R_{0,xx}}{R_{0,xx}}$ ($\frac{R_{\alpha,xy}-R_{0,xy}}{R_{0,xx}/7.53}$) in the longitudinal (transverse) direction and 7.53 is the geometric conversion factor. }
		\label{fig:SMR_data} 
	\end{figure}

Figure \ref{fig:SMR_data}(a) shows the relative change of resistivity and the according change of resistance ($0.25\times10^{-3}$ and $124~$m$\Omega$ for 8 T, respectively) of the longitudinal SMR at room temperature. The resistivity is minimal when the accumulated spins and the magnetic moments are colinear since the interface electrons are deflected by the ISHE into the direction of the current at the corresponding field angles, $0\degree$ and $180\degree$. At other field angles, there is a spin current into the NiO which decreases the spin accumulation and increases the resistivity as $\sin\alpha$. 

Futhermore, the spin transfer alters the spin direction of the accumulated electrons. This affects the direction of the ISHE mediated electron deflection, creating transverse deflection and more scattering of electrons in the longitudinal geometry. Since the change in spin direction is a function of its relative direction with the NiO magnetic moments, this also leads to an increase  in resistance as $\sin\alpha$. The combination of the two angular dependencies cause the observed modulation of the resistance $R_{\alpha,xx}\propto \sin^2\alpha$ in the longitudinal geometry. 

In FMs however, the longitudinal resistivity is maximal when the magnetic field and the spin accumulation are perpendicular since the magnetic moments coherently follow the applied magnetic field. Therefore, a phase shift of $90\degree$ arises in the angular dependence of the SMR of an AFM as compared to a FM. The $\cos^2\alpha$ angular dependent SMR of a FM\cite{Saitoh:2013} changes into $\cos^2(\alpha-90\degree)=\sin^2\alpha=1-\cos^2\alpha$ for an AFM. The modulation has changed from a positive to a negative $\cos^2\alpha$, giving reason to call it a negative SMR.

Since a transverse deflection creates a voltage difference in the transverse geometry as $-\cos\alpha$, the angular dependence in the transverse geometry is $R_{\alpha,xy} \propto -\cos\alpha\times\sin\alpha = -\frac{1}{2}\sin2\alpha$. The relative changes in resistivity and resistance ($0.24\times10^{-3}$ and $15 ~$m$\Omega$ for 8~T, respectively) are measured as shown in Fig. \ref{fig:SMR_data}(b). These transverse results agree with the longitudinal relative change of resistivity taken the geometric conversion factor of 7.53 into account, which is the ratio between length and width of the main Hall bar segment. The peak to peak changes of the angular dependent resistivity parts in Eqs. (\ref{rhol}) and (\ref{rhot}) match up to an average factor of $0.95\pm0.06$. This means that for both the transverse and the longitudinal case, $\Delta\rho_{1}$ is equal and there is no difference in spread as a result of domain formation or anisotropy. For the etched sample however, this ratio is $0.85\pm0.02$, showing that the first part of Eqs. (\ref{rhol}) and (\ref{rhot}) are significantly different in this sample. 

\begin{figure*}[t]
	\includegraphics[width=16cm]{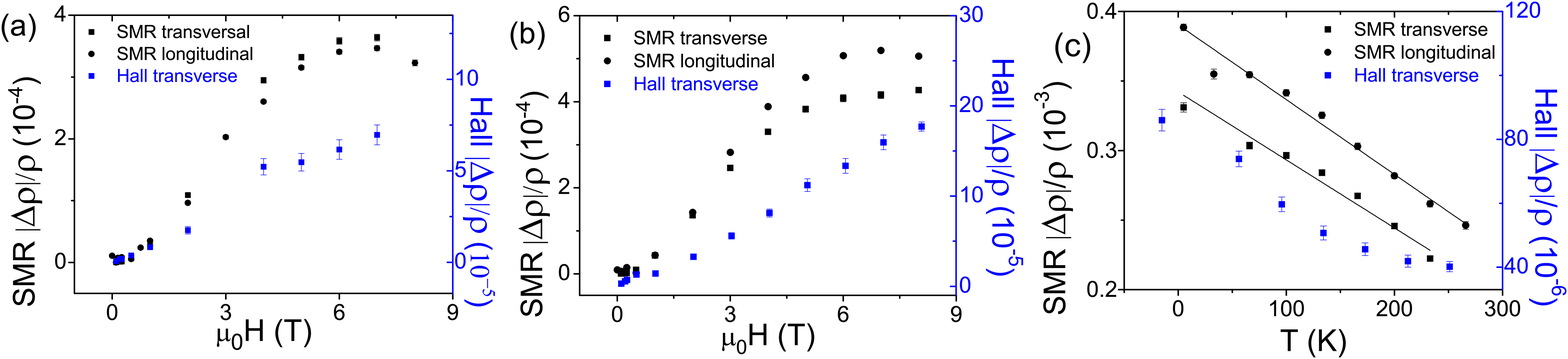}
	\caption{The relative SMR signal changes (black, left axis) and the transverse Hall contribution within this signal (blue, right axis) as a function of the magnetic field strength on the (a) unetched and (b) etched sample at 5~$\textrm{K}$. (c) The relative SMR signal as a function of temperature for the etched sample performed at 4~$\textrm{T}$ with fits shown as black lines.}
	\label{fig:SMR_data2}
	\end{figure*}

Besides the SMR signal, there is also a 360$\degree$ period Hall contribution in the transverse geometry which originates from a slight misalignment of the sample resulting in a small out-of-plane component of the magnetic field. A fit shows that the signal is one order of magnitude lower than the SMR signal, as can be seen in Fig. \ref{fig:SMR_data2}(a-c). The Hall component of both samples increases linearly with magnetic field strength as expected. However, the Hall contribution is $2.11\pm0.01$ times higher in the etched sample due to a larger misalignment angle. 
	
Figures \ref{fig:SMR_data2}(a) and \ref{fig:SMR_data2}(b) show relative signal strengths for the unetched and etched samples which increase with magnetic field strength and start to saturate around 6~T. The low SMR signal at low magnetic fields is attributed to the multidomain state of the single crystal under these conditions. A strong magnetic field could magnetoelastically increase the size of the domain with surface in the easy plane resulting in a higher SMR. The minimal magnetic field required for domain movement is about 0.24~T,\cite{Kurosawa:1980} while the transition to a single domain state occurs at fields of about 2.5~T in well oriented crystals.\cite{Roth:1960, RothSlack:1960, Slack:1960} In the obtained results there is an SMR signal starting from 0.25~T, although the saturation magnetic field does not match the 2.5~T reported. Patterning the NiO samples with Pt and Ti/Au contact leads might cause pinning of domain walls, what can be responsible for the higher fields needed for saturation.\cite{Roth:1960}

The background resistivity increases linearly with increasing temperature as expected for a metal. The relative SMR signal strength (Fig. \ref{fig:SMR_data2}(c)), however, decreases with temperature for the etched sample and is fitted with $\propto (T_N-T)^{0.7}$. This expression for the square of the order parameter, depending on $\propto(T_N-T)^{0.35}$, is obtained by a mean-field-like approach, including small fluctuations around the average magnetization.\cite{Chaikin:1995} The obtained parameter of the N\'eel temperatures are $551\pm16$~K and $531\pm25$~K for the longitudinal and transverse SMR signals, respectively, and compare relatively well to the established value of $523$~K. 

The spin-mixing conductance is $\sim 10^{18}~\Omega^{-1}$m$^{-2}$ as calculated from Eq. \ref{eq:SMR} for the unetched sample with $\lambda$, $d_N$ and $\theta$ assumed being constants of $1.1$~nm, $5$~nm and $0.08$ respectively.\cite{Vlietstra:2013} All fabrication steps of both samples are the same, except for the argon plasma treatment. Therefore, the magnitude of the SMR signal must originate from an enhancement of the spin-mixing conductance by a factor of two. So the etching significantly increases the interface transparancy. These spin-mixing conductance results are comparable to the SMR results of heterostructures with ferrimagnets such as Pt/YIG.\cite{Qiu:2013} Recently, an increase of the spin-mixing conductance after an etch step is observed in Pt/YIG systems as well, where it is further improved by an annealing step.\cite{Goennenwein:2017} \

Based on the bulk exchange coupling, the spin-mixing conductance is expected to be smaller than that of ferromagnets. A possible explanation for the high spin-mixing conductance in NiO is a lowered surface exchange coupling compared to the bulk.\cite{Xiao:2015} The results agree with the theoretical prediction of increased spin-mixing conductance due to interface disorder.\cite{Tserkovnyak:2014} The change in spin-mixing conductance due to the argon plasma treatment supports this theory as it affects the surface.\ 

An attempt to detect SMR in Pt/AFM bilayers has already been reported by Han $et$ $al$., where the SMR of Pt on bulk AFM SrMnO$_3$ had positive sign.\cite{Pan:2014} Taking our results into account, their study is an evidence that the SMR signals observed there are not due to an antiferromagnetic alignment but rather due to spin canting with respect to the magnetic field direction. 

In summary, SMR for  Pt/NiO(bulk) heterostructures with a (111) oriented surface has been studied. Negative longitudinal SMR is observed at fields higher than 0.25~T without the need of exchange coupling to an additional ferromagnet. The sign of the SMR indicates that the applied magnetic field couples perpendicular to the magnetic moments. The SMR signal increases by decreasing the temperature which is attributed to the lowering of the N\'eel order and its spatial fluctuations. The relative change of resistivity in the transverse geometry agrees with the longitudinal geometry in the sample without argon plasma treatment and follows $R_{\alpha,xy} \propto -\frac{1}{2}\sin2\alpha$ and $R_{\alpha,xx} \propto \sin^2\alpha$ angular dependence, respectively.\ 
A maximum SMR signal is observed around $6~$T with a corresponding spin-mixing conductance of $\sim10^{18}~\Omega^{-1}$m$^{-2}$ which is comparable to that of YIG. 
SMR proves to be an effective technique to investigate and manipulate the magnetic properties of AFMs. The simultaneous electrical injection and detection of spin currents, while having control over the magnetic moment directions, opens up the possibility of ultrafast and lossless AFM transport devices.

We acknowledge J. Baas, H. M. de Roosz, J. G. Holstein, H. Adema and T. J. Schouten for their technical assistance and appreciate M. V. Mostovoy, R. A. Duine and S. A. Bender for discussion. This work is part of the research program Magnon Spintronics (MSP) No. 159 financed by the Nederlandse Organisatie voor Wetenschappelijk Onderzoek (NWO). Further support by the DFG Priority Programme 1538 “Spin-Caloric Transport” (KU 3271/1-1) is gratefully acknowledged.
	
	

	\vspace{1cm}


\begin{thebibliography}{43}%
	
\bibitem{Barnas:1989}
{M.~N.~Baibich, J.~M.~Broto, A.~Fert, F.~Nguyen Van Dau, F.~Petroff, P.~Etienne, G.~Creuzet, A.~Friederich, and J.~Chazelas, Phys. Rev. Lett.~\textbf{61}, 2472 (1988).}

\bibitem{Binasch:1989}
{G.~Binasch, P.~Grünberg, F.~Saurenbach, W.~Zinn, Phys. Rev. B, \textbf{39} 4828 (1989).}

\bibitem{Julliere:1975}
{M.~Julliere, Phys. Lett. A,~\textbf{54}, 225 (1975).}

\bibitem{Wadley:2016}
{P.~Wadley, B.~Howells, J.~\v{Z}elezn\'{y}, C.~Andrews, V.~Hills, R.~P.~Campion, V.~Nov\'{a}k, K.~Olejn\'{i}k, F.~Maccherozzi, S.~S.~Dhesi, S.~Y.~Martin, T.~Wagner, J.~Wunderlich, F.~Freimuth, Y.~Mokrousov, J.~Kune\v{s}, J.~S.~Chauhan, M.~J.~Grzybowski, A.~W.~Rushforth, K.~W.~Edmonds, B.~L.~Gallagher, T.~Jungwirth, Science,~\textbf{351}, 6273 (2016).}

\bibitem{Cheng:2014}
{R.~Cheng, J.~Xiao, Q.~Niu, A.~Brataas,Phys. Rev. Lett. \textbf{113}, 057601 (2014).}

\bibitem{Tserkovnyak:2014}
{S.~Takei,1 B.~I.~Halperin, A.~Yacoby, T.~Tserkovnyak, Phys. Rev. B \textbf{90}, 094408 (2014).}

\bibitem{Xiao:2015}
{M.~W.~Daniels, W.~Guo, G.~M.~Stocks, D.~Xiao, J.~Xiao, New J. Phys. \textbf{17}, 103039 (2015).}

\bibitem{Rasing:2004}
{A.~V.~Kimel, A.~Kirilyuk, A.~Tsvetkov, R.~V.~Pisarev, T.~Rasing, T.  Nature \textbf{429}, 850–853 (2004).}

\bibitem{Brataas:2016}
{A.~Qaiumzadeh, H.~Skarsv\aa{}g, C.~Holmqvist, A.~Brataas, arXiv:1612.07440v1 (2016).}

\bibitem{Jungwirth:2016}
{T.~Jungwirth, X.~Marti, P.~Wadley, J.~Wunderlich, Nat. Nanotechnol.~\textbf{11}, 231 (2016).}

\bibitem{Kato:2004}
{Y.~K.~Kato, R.~R.~Myers, A.~C.~Gossard, D.~D.~Awschalom, Science \textbf{306},1910 (2004).}

\bibitem{Tatara:2006}
{E.~Saitoh, M.~Ueda, H.~Miyajima, G.~Tatara, Appl. Phys. Lett.~\textbf{88}, 182509 (2006).}

\bibitem{Saitoh:2013}
{H.~Nakayama, M.~Althammer, Y.~T.~Chen, K.~Uchida, Y.~Kajiwara, D.~Kikuchi, T.~Ohtani, S.~Gepr\"{a}gs, M.~Opel, S.~Takahashi, R.~Gross, G.~E.~W.~Bauer, S.~T.~B.~G\"{o}nnenwein, E.~Saitoh, Phys. Rev. Lett.~\textbf{110}, 206601 (2013).}

\bibitem{Bauer:2013}
{Y.~T.~Chen, S.~Takahashi, H.~Nakayama, M.~Althammer, S.~T.~B.~Goennenwein, E.~Saitoh, and G.~E.~W.~Bauer, Phys. Rev. B \textbf{87}, 144411 (2013).}

\bibitem{Vlietstra:2013}
{N.~Vlietstra, J.~Shan, V.~Castel, J.~Ben~Youssef, G.~E.~W.~Bauer, B.~J.~van~Wees, Appl. Phys. Lett. \textbf{103}, 032401 (2013).}

\bibitem{Goennenwein:2013}
{M.~Althammer, S.~Meyer, H.~Nakayama, M.~Schreier, S.~Altmannshofer, M.~Weiler, H.~Huebl, S.~Gepr\"{a}gs, M.~Opel, R.~Gross, D.~Meier, C.~Klewe, T.~Kuschel, J.~M.~Schmalhorst, G.~Reiss, L.~Shen, A.~Gupta, Y.~T.~Chen, G.~E.~W.~Bauer, E.~Saitoh, S.~T.~B.~G\"{o}nnenwein, Phys. Rev. B ~\textbf{87}, 224401 (2013).}

\bibitem{Ganzhorn:2016}
{K.~Ganzhorn, J.~Barker, R.~Schlitz, B.~A.~Piot, K.~Ollefs, F.~Guillou, F.~Wilhelm, A.~Rogalev, M.~Opel, M.~Althammer, S.~Gepr¨ags, H.~Huebl, R.~Gross, G.~E.W.~Bauer, S.~T.~B.~G\"{o}nnenwein, Phys. Rev. B ~\textbf{94}, 094401 (2016).}

\bibitem{Aqeel:2016}
{A.~Aqeel, N.~Vlietstra, A.~Roy, M.~Mostovoy, B.~J.~van~Wees, T.~T.~M.~Palstra, Phys. Rev. B~\textbf{94}, 134418 (2016).}

\bibitem{Kurosawa:1980}
{S.~Saito, M.~Miurat, K.~Kurosawa, J. Phys. C: Solid St. Phys. \textbf{13}, 1513 (1980).}

\bibitem{Slack:1960}
{G.~A.~Slack, J. Appl. Phys. \textbf{31}, 1571 (1960).}

\bibitem{Roth:1960}
{W.~L.~Roth, J. Appl. Phys. \textbf{30}, 2000 (1960).}

\bibitem{Yang:2014}
{H.~Wang , C.~Du , P.~C.~Hammel, F.~Yang , Phys. Rev. Lett.~\textbf{113} 097202 (2014).}

\bibitem{Kent:2016}
{Y.~M.~Hung, C.~Hahn, H.~Chang, M.~Wu, H.~Ohldag, A.~D.~Kent, AIP Adv.~\textbf{7}, 055903 (2017).}

\bibitem{QiuSaitoh:2016}
{Z.~Qiu, J.~Li, D.~Hou, E.~Arenholz, A.~T.~N\textsc{\char13}Diaye, A.~Tan, K.~Uchida, K.~Sato, S.~Okamoto, Y.~Tserkovnyak, Z.~Q.~Qiu, E.~Saitoh, Nat. Comm.~\textbf{7}, 12670 (2016).}

\bibitem{Tokura:2015}
{S.~Seki, T.~Ideue, M.~Kubota, Y.~Kozuka, R.~Takagi, M.~Nakamura, Y.~Kaneko, M.~Kawasaki, Y.~Tokura, Phys. Rev. Lett. \textbf{115}, 266601 (2015).}

\bibitem{Heremans:2016}
{A~Prakash, J.~Brangham, F.~Yang,J.~P.~Heremans, Phys. Rev. B~\textbf{94}, 014427 (2016).}

\bibitem{Chien:2016}
{W.~Lin, K.~Chen,S.~Zhang, C.~L.~Chien, Phys. Rev. Lett.~\textbf{116}, 186601 (2016).}

\bibitem{Chienelectrical:2016}
{W.~Lin, C.~L.~Chien, Phys. Rev. Lett.~\textbf{118}, 067202 (2017).}

\bibitem{Saitoh:2016}
{D.~Hou, Z.~Qiu, J.~Barker, K.~Sato, K.~Yamamoto, S.~U.~Velez, J.~M.~Gomez-Perez, L.~E.~Hueso, F.~Casanova, E.~Saitoh, Phys. Rev. Lett.~\textbf{118}, 147202 (2017).}

\bibitem{Shang:2016}
{T.~Shang, Q.~F.~Zhan, H.~L.~Yang, Z.~H.~Zuo, Y.~L.~Xie, L.~P.~Liu, S.~L.~Zhang, Y.~Zhang, H.~H.~Li, B.~M.~Wang, Y.~H.~Wu, S.~Zhang, R.-W. Li, Appl. Phys. Lett. \textbf{109}, 032410 (2016).}

\bibitem{Bauer:2000}
{A.~Brataas, Y.~V.~Nazarov G.~E.~W.~Bauer, Phys. Rev. Lett. \textbf{84}, 11 (2000).}

\bibitem{Aqeel:2014}
{A.~Aqeel, I.~J.~Vera-Marun, B.~J.~van~Wees, T.~T.~M.~Palstra, J. Appl. Phys. \textbf{116}, 153705 (2014).}

\bibitem{RothSlack:1960}
{W.~L.~Roth, G.~A.~Slack, J. Appl. Phys. \textbf{31}, S352 (1960).}

\bibitem{Chaikin:1995}
{P.~M.~Chaikin, T.~C.~Lubensky, \textit{Principles of Condensed Matter Physics} (Cambridge University Press, Cambridge, England 1995).}

\bibitem{Qiu:2013}
{Z.~Qiu, K.~Ando, K.~Uchida, Y.~Kajiwara, R.~Takahashi, H.~Nakayama, T.~An, Y.~Fujikawa, E.~Saitoh, Appl. Phys. Lett. \textbf{103}, 092404 (2013).}

\bibitem{Goennenwein:2017}
{S.~Putter, S.~Gepr\"{a}gs, R.~Schlitz, M.~Althammer, A.~Erb, R. Gross, S.~T.~B.~Goennenwein, Appl. Phys. Lett. \textbf{110}, 012403 (2017).}

\bibitem{Pan:2014}
{J.~H.~Han, C.~Song, F.~Li, Y.~Y.~Wang, G.~Y.~Wang, Q.~H.~Yang, F.~Pan, Phys. Rev. B~\textbf{90}, 144431 (2014).}

\end{thebibliography}
\end{document}